# A Solid-State System for High Temporal Resolution Fluorescence Lifetime Measurements

M. Gersbach, D. L. Boiko, C. Niclass, C. Petersen and E. Charbon

Ecole Polytechnique Fédérale de Lausanne (EPFL), 1015 Lausanne, Switzerland

**Abstract**

A system for time-correlated single photon counting applications based on CMOS single photon avalanche diodes is presented. An instrument response function of 79 ps allows a fluorescence decay of high-affinity $Ca^{2+}$ indicator Oregon Green BAPTA-1 to be analyzed with unprecedented temporal resolution in the two-photon excitation regime. A triple exponential decay model is shown to best fit the fluorescence dynamics of OGB-1.

Fluorescence Lifetime Imaging Microscopy (FLIM) is used to locally probe the chemical environment of fluorophores, e.g., ion concentration, pH or oxygen content [1,2]. Several approaches were proposed to acquire time-resolved fluorescence images, among which a widely used technique is Time-Correlated Single Photon Counting (TCSPC) [3]. Thanks to detectors exhibiting single-photon sensitivity and high temporal resolution it is possible to measure the probability distribution of photon arrival times with very high accuracy, while remaining insensitive to instabilities of the excitation beam intensity. For TCSPC applications, the sensors of choice today are Photomultiplying Tubes (PMTs) and Microchannel-Plates (MCPs). These devices exhibit single photon sensitivity and low timing jitter, but are bulky and require high operating voltages. As an alternative to PMTs, Single Photon Avalanche Diodes (SPADs) have been proposed for their temporal resolution and sensitivity [4-6].

In this letter, we present a system for TCSPC applications based on SPADs implemented in CMOS technology. Our SPADs [7,8] are integrated in a 32x32 array and incorporate on-chip high bandwidth I/O circuitry. Using a single SPAD of the array, we demonstrate an Instrument Response Function (IRF) of 79 ps and test the efficiency of our TCSPC system by measuring the fluorescence decay of the high-affinity $Ca^{2+}$ indicator Oregon Green BAPTA-1 (OGB-1) under two-photon excitation conditions. Our measurements reveal a previously unobserved fast dynamic behavior of OGB-1 fluorescence. We provide a model and a detailed comparison with previously reported data [1,4,11], which are acquired using a PMT with temporal resolution of ~200ps.

Our SPAD-based [7,8] TCSPC system is depicted in Fig. 1. The active region of a SPAD [Fig.1(b)] consists of a $p^+$–n diode biased above its breakdown voltage, thus operating in the Geiger mode. When a photon is absorbed in the multiplication region, an avalanche is triggered. The avalanche breakdown is subsequently quenched by an on-chip ballast resistor, which is used to read out the photo-detection events. The single photon detection probability is 25 % at 500 nm wavelength [7-8]. At room temperature, due to a small active region (7 μm diameter), our SPADs show an extremely low dark count rate (<10 Hz). The dead time is 25 ns with negligible afterpulsing (<0.1 %).

Fluorescent molecules are excited in two-photon regime [9], using a mode-locked Ti:Sapphire laser (MaiTai, Spectra Physics) emitting 100 fs optical pulses at 800 nm wavelength [Fig.1(a)]. The excitation beam is focused on a sample using a 20x microscope objective (XLUMPlanFL, Olympus), which also serves to collect the fluorescent emission of OGB-1 molecules treated here. The fluorescence emission centered at 520 nm wavelength is directed towards the SPAD by a dichroic beam splitter (DBS) and a filter (E650SP, Chroma Technology) that suppress backscattered excitation pulses. Another 20x objective images the emission spot onto the SPAD. The low quantum yield of the two-photon excitation fluorescence and mismatch between the SPAD sensitive area (40 $\mu m^2$) and the fluorescent spot image result in a count rate at the detector of ~10 kHz (at 9 mW of excitation power at the sample and 80 MHz repetition rate of excitation pulses). The time discrimination is performed in a "reversed start-stop" configuration. The start signal is a photo-detection event at the SPAD while the emission of successive excitation pulse from Ti:Sapphire laser is used as a stop signal. For precise timing of photon arrivals, we use a 6 GHz bandwidth oscilloscope (WaveMaster, LeCroy) incorporating a time-to-digital-converter (TDC) and enabling to compute histograms of photon arrivals.

Fig. 2 (inset) shows the IRF histogram of the entire system (black curve). It is recorded by measuring, at 400 nm wavelength, the hyper-Rayleigh scattering of femtosecond pulses (800 nm wavelength) in a solution of colloidal gold particles (G1652, Sigma-Aldrich) [10]. At the incident power of 90 mW, the average count rate of the detector is just 600 Hz. The measured photon arrival time jitter of 79 ps (FWHM) is dominated by the SPADs time-response characteristics. The IRF is only slightly asymmetrical and assumes a Gaussian-curve approximation $IRF(t) \propto \exp(-t^2/2\sigma_{IRF}^2)$.

The FLIM-TCSPC system was tested using fluorescent samples composed of 2 µl OGB-1 dye (O6806, Molecular probes) and 20 µl calcium buffer solutions from the calibration kit (C3008MP, Molecular Probes) with quoted free $Ca^{2+}$ concentrations in the range of 17 nM to 39 µM. Several measured fluorescence decay curves in the two-photon excitation regime are shown in Fig. 2. The corresponding fluorescence lifetimes were obtained from the numerical analysis of these data.

Thorough numerical fit must take into account the IRF of the system [11]. As opposed to conventional systems with strongly asymmetric IRF, the Gaussian-like IRF of our system assumes a simple analytical expression for the measured fluorescence decay. Thus, for a train of excitation pulses of period $T$, each term of a multi-exponential decay process reads

$$I(t)_j = \frac{1}{2}\left[\frac{e^{T/\tau_j}+1}{e^{T/\tau_j}-1} - \operatorname{erf}\left(\frac{\sigma_{IRF}}{\tau_j\sqrt{2}} - \frac{t}{\sigma_{IRF}\sqrt{2}}\right)\right]\exp\left(-\frac{t}{\tau_j} + \frac{\sigma_{IRF}^2}{2\tau_j^2}\right) \qquad j=[f,i,s], \quad (1)$$

where a triple-exponential decay is assumed and index $j$ indicates fast, intermediate and slow temporal components, $\tau_j$ is the fluorescence emission lifetime ($\tau_f < \tau_i < \tau_s$) and $\operatorname{erf}(z) = \frac{2}{\sqrt{\pi}}\int_0^z \exp(-\xi^2)d\xi$ is the error function. Eq. (1) takes the periodic train of excitation pulses and response time jitter of the SPAD into account, such that our data in Fig. 2 does not require deconvolution processing.

Analysis of the OGB-1 fluorescence reveals the best agreement with a triple exponential decay approximation. The data in Fig. 2 are modeled using the function $I_f A_f + I_i A_i + I_s (1 - A_f - A_i)$ with the fast $I_f(t)$, intermediate $I_i(t)$ and slow $I_s(t)$ decaying components (1) of (normalized) partial intensities $A_f$, $A_i$ and $A_s = 1 - A_f - A_i$, respectively. Fig. 3 details a comparison between the double- and triple-exponential models applied to fluorescence from a 38 nM $Ca^{2+}$ concentration sample. In both cases, the residues [bottom panels] do not show any bias that might be caused by the Gaussian curve approximation of the IRF. However, the accuracy of the numerical fit is improved in case of the triple-exponential model. Wilms *et al.* [11] have made the same observation for OGB-1 fluorescence in the absence of $Ca^{2+}$. However, it was attributed to contaminating dye derivatives and a double-exponential model has been utilized for samples containing $Ca^{2+}$. Previously reported data [1,4,11] for OGB-1 fluorescence lifetimes thus assume the double-exponential decay approximation. We argue that PMT-based systems used so far exhibit insufficient temporal resolution to enable observation of the triple-exponential behavior.

Using our triple-exponential decay model (1) and assuming that the lifetimes are independent of the calcium buffer [1,11], we applied a global numerical analysis to our data, yielding the lifetimes $\tau_f$, $\tau_i$, and $\tau_s$ and the partial intensities $A_f$, $A_i$ and $A_s$ of OGB-1 fluorescence in function of free $Ca^{2+}$ concentration (Fig. 4). Temporal resolution of our system allows the short lifetime component ($\tau_f$~190ps) to be unambiguously resolved in the background of the intermediate- and long-living fluorescence ($\tau_i$~770ps and $\tau_s$~4.2ns).

The partial intensities $A_f$ and $A_i$ decrease rapidly while the slow-component intensity $A_s = 1 - A_f - A_i$ increases with $Ca^{2+}$ concentration (top panel). We attribute, as usual, the short- ($\tau_f$) and long-lifetime ($\tau_s$) components respectively to $Ca^{2+}$-free OGB-1 and to the $Ca^{2+}$-bound OGB-1 complex with a covalent bond and a simple 1:1 molar ratio. The intermediate lifetime component $\tau_i$ can then be attributed to the $Ca^{2+}$- OGB-1 complex with coordinate covalent bonds and a molar ratio of 1:n. Assuming that the intensities are proportional to the concentration of corresponding OGB-1 complexes, one can easily find a relationship between the free calcium concentration $[Ca^{2+}]$ and the dissociation constant $K_D$ of the main (1:1) complex:

$$[Ca^{2+}] = K_D \frac{A_s}{1-A_s}\left[1+\left(\frac{A_f}{A_i}\right)^{-1}\right], \quad \frac{A_f}{A_i} \propto [Ca^{2+}]^{-1/n} \tag{2}$$

The intensity ratio $A_f/A_i$ of the short and intermediate lifetime components decays with $Ca^{2+}$ concentration (bottom panel) and can be used for fast calcium sensing. This opens the way to increased data acquisition speed in real-time imaging applications. The numerical fit reports a value n=4.7 indicating that a coordinate complex with 1:5 molar ratio is dominating the $\tau_i$ component [*e.g.* $(OGB-1)_5H_{28}Ca$]. The dissociation constant $K_D$ reported by the fit is 180 nM, in agreement with the value quoted by the manufacturer (170 nM).

For the long lifetime component $A_s$, Fig. 4 indicates good agreement with the data published by Agronskaia *et al.* [4] and Wilms *et al.* [11]. In [1], Lakowicz reported $\tau_s$ of 4 ns in $Ca^{2+}$ saturated solution, which well agrees with our data. In [4], $\tau_s$ varies in the range 2.6-3.7 ns. In [11], $\tau_s$=3.63 ns while $A_s$ is less than one in $Ca^{2+}$-saturated buffer, which is again attributed to dye impurities; $K_D$ of ~300 nM is considered as an apparent value and, to meet the specification, it is corrected by allowing for a difference in two-photon *absorption* cross-sections of Ca-bound and Ca-free OGB-1 molecules.

For unbound OGB-1, the articles cited above report a single lifetime component of 700 [1], 290-420 [4] and 346 ps [11], respectively. These results correspond to the combined effect of the decay processes with the lifetimes $\tau_f$ and $\tau_i$ in our measurements. Because of the short lifetime $\tau_f$ of 190 ps, we believe that this component may not have been resolved with certitude in previous studies. The experimental setup in [11] relies on a commercially available PMT with a quoted timing jitter of 200 ps. Other data [1,4] are reported without timing resolution of experimental setups.

In summary, the OGB-1 fluorescence lifetime study by using novel single-photon avalanche detectors proves that CMOS SPAD array technology is well suited for TCSPC applications. As opposed to previous reports, a triple-exponential fluorescence decay is observed for OGB-1 in Ca buffer solution allowing its dissociation constant to be measured directly. Furthermore, the use of standard CMOS technology opens the way to the integration of photon arrival timing measurements and data processing on chip.

This work was partially funded by the Swiss National Science Foundation and by Centre SI of EPFL.

**References**

[1]  J. Lakowicz, *Principles of Fluorescence Spectroscopy*, Plenum, New York (1999)

[2]  K. Suhling, P.M.W French and D. Phillips, "Time-resolved fluorescence microscopy," *Photochem. Photobiol. Sci.* **4** (1), 13-22 (2005)

[3]  D.V. O'Connor and D. Phillips, *Time-correlated Single Photon Counting*, Academic Press, London (1984)

[4]  S. Cova, A. Longoni, A. Andreoni and R. Cubeddu, "A Semiconductor Detector for Measuring Ultraweak Fluorescence Decays with 70 ps FWHM Resolution," *IEEE J. Quantum Electron.* **19** (4), 630-634 (1983)

[5]  A. Gulinatti, P. Maccagnani, I. Rech, M. Ghioni and S. Cova, "35ps time resolution at room temperature with large area single photon avalanche diodes," *Electron. Lett.* **41** (5), 272-274 (2005)

[6]  M. Gersbach, C. Niclass, M. Sergio, D. L. Boiko, C. Petersenand E. Charbon, "Time-Correlated Fluorescence Microscopy Using a Room Temperature Solid-State Single Photon Sensor", *9th International conference on near-field optics, nanophotonics and related techniques (NFO-9),* TuP-78 (2006)

[7]  C. Niclass, A. Rochas, P.A. Besse, E. Charbon, "A CMOS Single Photon Avalanche Diode Array for 3D Imaging", *IEEE International Solid-State Circuits Conference (ISSCC)*, 120-121 (2004)

[8]  C. Niclass, A. Rochas, P.A. Besse, and E. Charbon, "Design and Characterization of a CMOS 3-D Image Sensor Based on Single Photon Avalanche Diodes", *IEEE J. Solid-State Circuits* **40** (9), 1847-1854 (2005).

[9] W. Denk, J.H. Strickler and W.W. Webb, "Two-photon laser scanning fluorescence microscopy," *Science* **248** (4951), 73-76 (1990)

[10] A. Habenicht, J. Hjelm, E. Mukhtar, F. Bergstrom and L.B.A. Johansson, "Two-photon excitation and time-resolved fluorescence: 1. The proper response function for analyzing single-photon counting experiments," *Chem. Phys. Lett.* **354** (5-6), 367-375 (2002)

[11] C.D. Wilms, H. Schmidt and J. Eilers, "Quantitative two-photon $Ca^{2+}$ imaging via fluorescence lifetime analysis," *Cell Calcium* **40** (1), 73-79 (2006)

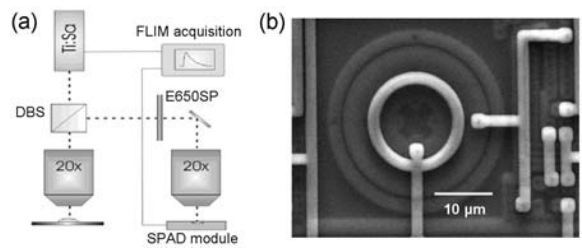

**Fig. 1.** (a) Schematic of the experimental setup for fluorescence lifetime measurements. (b) Scanning electron microscope image of a SPAD.

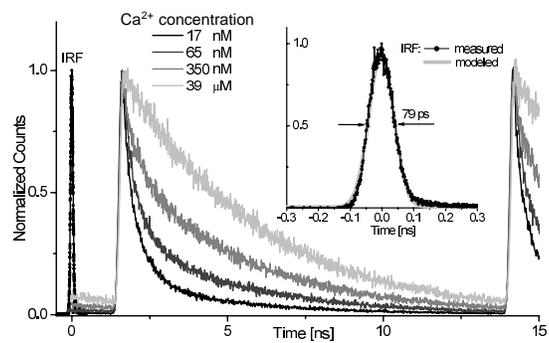

**Fig. 2.** Measured fluorescence decay of OGB-1 dye molecules in the presence of different calcium concentrations. The inset shows time-resolved hyper-Rayleigh scattering from colloidal gold particles used to measure the IRF.

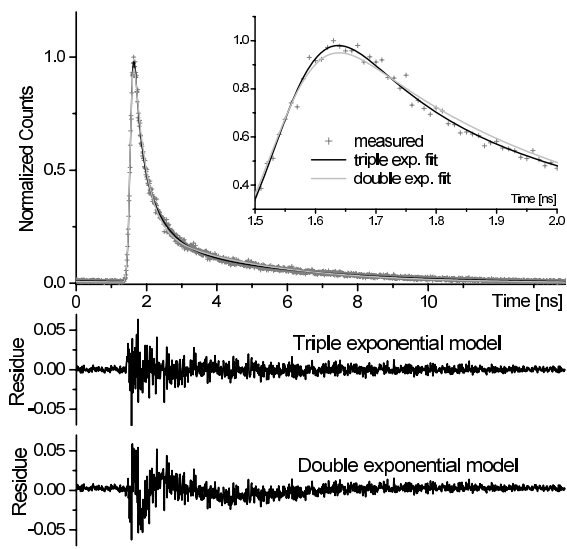

**Fig. 3.** Fluorescence decay of OGB-1 in a 38 nM free $Ca^{2+}$ concentration buffer. Top panel: measured (points) and numerically fitted curves using the double- (gray color) and triple-(black color) exponential decay models. The inset shows a close-up of the initial interval of 500ps width. Bottom panels: the residues for the two models.

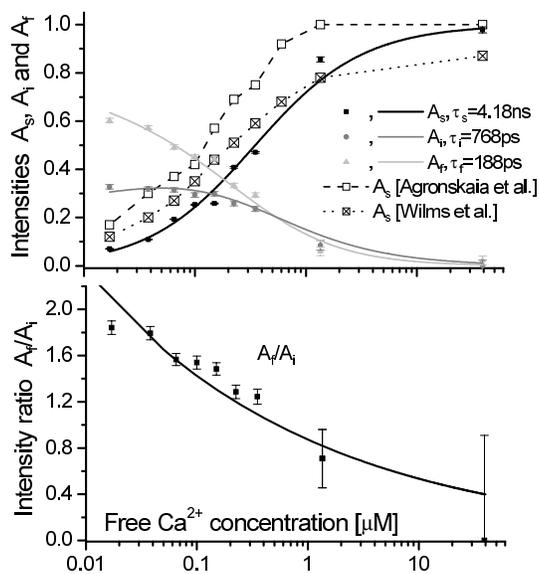

**Fig. 4.** Partial intensities of the slow ($A_s$), intermediate ($A_i$) and fast ($A_f$) decay components of OGB-1 fluorescence (top panel) and the ratio $A_f/A_i$ (bottom panel) as a function of the free $Ca^{2+}$ concentration. (For $Ca^{2+}$-saturated solution, both $A_f$ and $A_i$ approach null.) The dark solid lines show the fit with Eq.(2) yielding $K_D$=182±2 nM. The lifetimes are 4.18±0.01 ns, 768±16 ps and 188±6 ps. Our data is shown in comparison with the data from [4] and [11].